\documentclass[12pt]{article}

\usepackage{amsmath,amssymb,amsthm,eurosym,lineno,mathrsfs}
\RequirePackage[colorlinks=true,allcolors=blue]{hyperref}

\numberwithin{equation}{section}

\begin{document}


\title{\bf\large Return Optimization Securities and Other Remarkable Structured Investment Products: Indicators of Future Outcomes for U.S. Treasuries?}

\author{{Donald St. P. Richards}\thanks{Department of Statistics, 
Penn State University, University Park, PA 16802.
\endgraf
\endgraf
}
}

\date{March 27, 2013}
\maketitle

\begin{abstract}
\footnotesize{
We analyze four structured products that have caused severe losses to investors in recent years.  These products are: return optimization securities, yield magnet notes, reverse exchangeable securities, and principal-protected notes.  We describe the basic structure of these products, analyze them probabilistically using the Law of Total Expectation, and assess the practical implications of buying them in the mid-2000s.  By estimating expected rates of return under various scenarios, we conclude in each case that buyers were likely to experience grave difficulties from the start.  By inspecting various prospectuses, we detect that many structured products were designed to the detriment of buyers and to the advantage of the issuing banks and broker-dealers.  Therefore, we find it difficult to understand why any investment advisor, in exercising fiduciary care of clients' funds, would have advised a client to purchases these products in the mid-2000's.  

In light of these results, we fear that the on-going worldwide financial crisis will be lengthened because of these structured products and others even more arcane than the ones considered here.  We note that problems caused by structured products have increased investors' fears about the economy and the financial markets, causing many of them to purchase U.S. Treasury securities at negative real-interest rates.  This has caused increases in the prices of U.S. Treasuries to record levels, and we fear for the day when this trend reverses.
}
\end{abstract}


\parskip=2.5pt

\section{Introduction}
\label{intro}
\setcounter{equation}{0}

In a recent article, Richards and Hundal \cite{richardshundal} analyzed a class of structured financial products, called constant proportion debt obligations, which became insolvent rapidly and caused losses worldwide in 2006--2008.  In the present paper, we analyze four structured products devised by prominent investment banks and sold in large volumes to investors.  As with constant proportion debt obligations, our attention was drawn to these products by news articles describing the woes of investors in those products.  

The structured products we analyze are: return optimization securities, yield magnet notes, reverse exchangeable securities, and principal-protected notes.  We describe the basic structure of these products, analyze them probabilistically, and consider the practical implications of buying them in the mid-2000s.  We conclude in each case that buyers were likely to experience difficulties from the start.  

Moreover, we find it difficult to believe that a typical investment adviser was able to understand fully these financial products, so we conclude that investment advisers were incapable of exercising fiduciary duty to clients who purchased these products.  

As regards principal-protected notes and return optimization securities, it was an article by Papini \cite{papini} that drew our attention to those products.  Papini described the case of an investor who lost her entire investments in ``a \$225,000 guaranteed principal protected note and a \$75,000 reverse optimization note.''  The investor's lawyer ``argued that the notes were `speculative derivative securities' and were `unsuitable' for unsophisticated investors.''  The investor ultimately was partially successful at recovery through arbitration by a panel of the Financial Industry Regulation Authority (FINRA); but more ominous was the lawyer's comment that there were ``many similar pending cases,'' indicating that large numbers of similar investors had lost their investments.   

On reading Papini's article, we wondered whether a typical investor or investment advisor could comprehend fully the risks of a return optimization security.  Having not heard of these structured products before reading that article, we decided to analyze them so as to understand the reasons for the investor's losses.  

The word ``optimization'' in the name of the product struck in our mind a clear warning bell.  The word ``optimization,'' while familiar to mathematicians, is not commonly found in the vocabulary of the typical small investor, and we detected immediately that financial engineers had been at work in the creation of ``return optimization securities,'' whatever they might be.  Indeed, an Internet search for the phrase ``return optimization'' reveals that it was Markowitz \cite{markowitz} who developed mathematical methods for solving the ``risk-return optimization'' problem.  

In studying those four structured products, we conclude that none are suitable for the typical unsophisticated investor.  We have a strong impression that these products were designed so as to place buyers at a disadvantage to the investment banks that originated the products and to the broker-dealers that sold them; moreover, the same impression applies to structured products as a class.  Therefore, we recommend that small investors regard structured products generally with great skepticism.  

A basic motivation for the purchasing of structured products is a desire for higher yield on fixed-income investments during a protracted period of low interest rates.  These low rates have enabled financial firms to engage in unprecedented speculative activities, which have then led to widespread uneasiness about the financial markets and the economy.  Investors, in an effort to assuage their anxieties, have purchased U.S. Treasury securities at record, bubble-like prices equivalent to negative interest rates.  To the extent that such a bubble exists, we truly fear for the day when it bursts.

\section{Return optimization securities}
\label{returnoptimizationsecurities}
\setcounter{equation}{0}

The nature of return optimization securities can be revealed by searching the website of the Securities and Exchange Commission (SEC) for the phrase ``return optimization'' and then downloading one of the many corresponding prospectuses.  

A prospectus for a ``Return Optimization Note with Partial Protection Linked to the S\&P 500 Index'' shows that these notes were typically sold for a price of \$10 per note with a term, or maturity period, of 369 days.  

Skepticism is heightened immediately by the maturity term of 369 days since a choice of 365 or 366 days, say, would have been more typical.  Despite much research on this issue, we could find no explanation for the choice of 369 days, and we wonder if the 369-day term was chosen to provide favorable tax treatment to the seller at the expense of the buyer.  Given the strange name, ``return optimization securities,'' and the curious maturity term of 369 days, a cautious investor should be nervous of these securities and should decline them.  

Denote by $I_0$ the price of the Standard \& Poor (S\&P) 500 index at the beginning of the maturity period, and denote by $I_1$ the price of the S\&P 500 index at the end of the maturity period.  Then the percentage return on the index over the maturity period is 
\begin{equation}
\label{spreturn}
I = \frac{I_1-I_0}{I_0}
\end{equation}

The net payment to the investor on the maturity date of a return optimization security depends on a mysterious number, which we denote by $M$.  Despite a close inspection of the prospectus, we were unable to ascertain how exactly the number $M$ was to be calculated; however, it is stated in the prospectus that 
\begin{equation}
\label{mystery}
0.25 \le M \le 0.30
\end{equation}

The total payment to investors at maturity is stated in the prospectus in a complicated manner, so much so that we find it difficult to believe that a small investor could extract or understand the crucial details.  We determined eventually that the net payment at maturity is given by the formula
\begin{equation}
\label{rosnet}
\hbox{Net payment } = \begin{cases}
\min(50I,10M), & \hbox{if } I \ge 0 \\
10I, & \hbox{if } I < 0
\end{cases}
\end{equation}
This formula is interesting.  Suppose that the S\&P 500 index were to decline 42\% during the 369-day maturity period; then $I = -0.42$ and, by (\ref{rosnet}), the investor's net payment is \$-4.2.  Simply put, an investor in return optimization securities would have lost 42\% of their capital in that 369-day period.  

Now let us consider the actual returns to an investor in a return optimization security over the 369-day period October 05, 2007 -- October 10, 2008, a period which exhibited events that were unprecedented in the history of the United States.  The following data were obtained from Google Finance: 

\begin{table}[!ht]
\caption{The closing value of the S\&P 500 index over a 369-day period}
\begin{center}
\renewcommand{\arraystretch}{1.10}
\begin{tabular}{|lcl|}
\hline
Date & \phantom{ABC} & Closing value \\
\hline
October 05, 2007 & & 1557.59 \\
October 10, 2008 & & \phantom{1}899.22 \\
\hline
\end{tabular}
\end{center}
\label{probtable}
\end{table}

\noindent
Then, the return on the index over the period October 5, 2007 - October 10, 2008 was 
$$
\frac{899.22 - 1557.59}{1557.59} = -0.42
$$
The decline of 42\% in the index in that period means that an owner of this return optimization security had permanent losses of 42\% of their capital.  Since the securities cannot discern their owners' level of sophistication then we must issue a warning:  {\it Return optimization securities may also be unsuitable for sophisticated investors!}  

To provide a balanced assessment of the securities, let us also determine how holders of the securities would have fared had the S\&P 500 index advanced 42\% during the maturity period.  By (\ref{rosnet}), the net payment is at most $10M$ if $I = 0.42$.  By (\ref{mystery}), $0.25 \le M \le 0.30$, so if the index advances 42\% then the investor's return is at most 30\%.  Hence, the buyer of a return optimization security is risking a 100\% loss vs. a gain of at most 30\%.  However, as any investor should know: {\it It is unwise to risk a 100\% loss in return for a small chance of at most a 30\% gain}.  

At this stage, it is hard to escape the conclusion that return optimization securities are incredibly risky for investors, perhaps even predatory in nature.  

To discover other dangerous aspects of these securities, we shall apply some elementary undergraduate-level probability theory.  Denote by $R$ the percentage net return on the investor's \$10 capital; by (\ref{rosnet}), we have 
\begin{equation}
\label{rospercentagenet}
R = \frac{\hbox{Net payment}}{10} = 
\begin{cases}
\min(5I,M), & \hbox{if } I \ge 0 \\
I, & \hbox{if } I < 0
\end{cases}
\end{equation}
and it follows from (\ref{rospercentagenet}) that 
\begin{equation}
\label{rosreturnupperbound}
R \ \begin{cases}
\le M, & \hbox{if } I \ge 0 \\
= I, & \hbox{if } I < 0
\end{cases}
\end{equation}
Denote by $P(I < 0)$ the probability that the index will provide negative return over the maturity period; then $P(I \ge 0) = 1 - P(I < 0)$ is the complementary probability.  Also, denote by $E(I \, | \, I < 0)$ the conditional expected value of the index {\it given} that the index provides a negative return over the maturity period.  

On applying to (\ref{rosreturnupperbound}) the Law of Total Expectation (see Ross \cite{ross}, p. 333) we obtain an upper bound for $E(R)$, the unconditional expected value of $R$:
\begin{equation}
\label{rosexpectedvaluebound}
E(R) \le M \cdot P(I \ge 0) + E(I \, | \, I < 0) \cdot P(I < 0). 
\end{equation}
It is obvious that 
$$
E(I \, | \, I < 0) \le 0,
$$
so it follows from (\ref{rosexpectedvaluebound}) that 
\begin{equation}
\label{rosexpectedvalueboundnegative}
E(R) \le M \cdot P(I \ge 0).
\end{equation}
We shall now show that formulas (\ref{rosexpectedvaluebound}) and (\ref{rosexpectedvalueboundnegative}) each have highly negative implications for holders of return optimization securities.  

To understand what can be discerned from these two formulas, some historical remarks are in order.  Recall that return optimization securities were sold in large quantities in 2006--2008 when the S\&P 500 index was near record levels; hence, both $I$ and $R$ were likely to be negative.  As I had argued in my \href{http://www.stat.psu.edu/~richards/talks/}{\textcolor{blue}{2006 Bowling Green Lectures}} \cite{richardsbgsu}, the financial markets were overdue for substantial declines and it was unlikely that the index would continue to rise.  

Suppose we had estimated in 2007 that there was a 10\% chance that the return on the index would be nonnegative over the ensuing year, i.e., $P(I \ge 0) = 0.1$.  By (\ref{rosexpectedvalueboundnegative}) and (\ref{mystery}), we have 
$$
E(R) \le 0.3 \cdot 0.1 = 0.03
$$
Hence, we would have predicted that investors holding return optimization securities in 2007 would obtain, on average, a return of {\it at most} 3\%.  Although this upper bound is positive, this should have been of little comfort to investors since many of them likely would obtain returns that were {\it negative} and substantially below the 3\% upper bound.  

Imagine that we had adopted a bearish view of the markets in 2007.  With the S\&P 500 index close to record territory then, if it were to decline, it could fall by at least 20\%; consequently, we estimate that $E(I \, | \, I < 0) \le -0.2$.  Suppose, as before, that $P(I \ge 0) = 0.1$, so that $P(I < 0) = 0.9$.  By (\ref{rosexpectedvaluebound}) and (\ref{mystery}), we obtain 
$$
E(R) \le (0.3 \cdot 0.1) + (-0.2 \cdot 0.9) = -0.15
$$
Under this scenario, we would have predicted in 2007 that investors who owned return optimization securities would have had average losses of {\it at least} 15\%.  

In a doomsday scenario, we had estimated in 2006 that if the S\&P 500 index were to decline then it would fall by more than 30\% (see Richards \cite{richardsbgsu}).  Under such a scenario, we have $E(I \, | \, I < 0) \le -0.3$.  Supposing again that  $P(I \ge 0) = 0.1$, so that $P(I < 0) = 0.9$, we obtain from (\ref{rosexpectedvaluebound}) and (\ref{mystery}) the inequality 
$$
E(R) \le (0.3 \cdot 0.1) + (-0.3 \cdot 0.9) = -0.24
$$
Therefore, the doomsday scenario would have led us to predict in 2007 that return optimization securities would generate average losses of {\it at least} 24\%.

\subsection*{Summary on return optimization securities}
\label{rosconclusions}

In studying the terms in (\ref{rosexpectedvaluebound}) and (\ref{rosexpectedvalueboundnegative}), we used simple bounds to infer that the expected return, $E(R) \le 0$.  We can use historical data on the S\&P 500 index to estimate all terms in those formulas and, when we do so, we arrive at the conclusion that $E(R)$ was substantially negative under a variety of plausible scenarios.  Therefore, we have found overwhelming evidence that buyers of return optimization securities in 2007--2008 were doomed from the outset to suffer massive losses.  

It also was ominous from the start that return optimization securities limited net returns to holders to {\it at most} 30\% but provided no limits on losses.  Such an arrangement allowed the sellers of the securities to profit if the equities markets advanced but provided no protection to buyers if markets declined.  In effect, buyers of return optimization securities insured the banks against substantial declines in the markets; such an arrangement allowed the banks to avoid direct stock sales on the open market, which could have triggered widespread market declines.  

Therefore, to purchase a return optimization security was to pay the banks for the privilege of insuring them against market declines; this is in analogy with a situation in which an insurance company pays a policyholder for the privilege of providing the insurance.  As a consequence, owners of return optimization securities paid thrice: first, when they purchased the security; second, when they realized capital losses as markets declined; and third, with the grief and woes caused by their losses.  

There are many variants of the return optimization securities that we have analyzed here.  From the perspective of an unsophisticated investor, these securities have names with strange terms such as ``buffered returns,'' ``contingent absolute return autocallable,'' ``contingent protection linked to [a] constant maturity commodity index.''  We have also found other types of return optimization securities that provide leveraged returns based on an equities or a commodities index.\footnote{Warning: Speculation in commodities is, for small investors, sheer madness!}  These securities all can be analyzed just as we have done above.  On the basis of a basic probabilistic analysis and our comments about strangely-named financial products we believe that, even today, these products entail enormous risks and we urge all investors to be skeptical about them.  

\medskip

To conclude, we find it difficult to ascertain why any investment adviser, in exercising {\it fiduciary care} of clients' funds, would have advised a client to purchase a return optimization security during the period 2006--2008.

\section{Yield magnet notes}
\label{yieldmagnetnotes}
\setcounter{equation}{0}

Here again, we encounter a structured product with a name which is unconventional for unsophisticated investors.  To understand its nature, we find at the SEC's website numerous prospectuses for yield magnet notes.  One such prospectus, chosen at random, 
describes yield magnet notes based on the stock prices of the following 15 companies chosen from the Dow Jones ``Global Titans'' Index: 

\begin{verbatim}
  1. American International Group 
  2. British Petroleum 
  3. Cisco Systems Inc. 
  4. Citigroup Inc.
  5. Dell Inc.
  6. Exxon Mobil Corporation
  7. General Electric Company
  8. GlaxoSmithKline plc
  9. Intel Corporation
 10. Johnson \& Johnson
 11. Microsoft Corporation
 12. Siemens AG
 13. Toyota Motor Corporation
 14. Vodafone Group plc 
 15. Wal-Mart Corporation
\end{verbatim}

\bigskip

The redemption price of the note is \$1,000.  The settlement date is March 15, 2006; the redemption date is March 15, 2011, a five-year period; and interest on the notes is paid annually on March 15 (the ``payment date'').

The coupon rate was calculated in an interesting way.  First, a ``determination date,'' or D-date, is fixed at three business days before March 15 of each year.  

\begin{table}[!h]
\caption{Calculation of the Coupon Rate for Yield Magnet Notes}
\begin{center}
\renewcommand{\arraystretch}{1.10}
\begin{tabular}{|lcl|}
\hline
Period & \phantom{ABC} & Coupon Rate \\
\hline
March 15, 2006 -- March 15, 2007 & & 5.5\% fixed rate \\
March 15, 2007 -- March 15, 2011 & & Variable rate \\
\hline
\end{tabular}
\end{center}
\label{yieldmagnetcouponrate}
\end{table}

\bigskip

The coupon rate for the period March 15, 2007 -- March 15, 2008 is computed as follows:

\bigskip

\noindent
Step 1: For the $i$th Titan stock, $i = 1,\ldots,15$, calculate 
\begin{equation}
\label{deltai}
\delta_i = \frac{\hbox{Stock price on D-date -- Stock price on
March 15, 2007}}{\hbox{Stock price on March 15, 2007}},
\end{equation}
which is the percentage return from the $i$th stock over the indicated period.  

\medskip

\noindent
Step 2: Calculate, for $i=1,\ldots,15$, 
\begin{equation}
\label{thetai}
\theta_i =
\begin{cases}
-0.125,   & \hbox{if } \ \delta_i < -0.125 \\
\delta_i, & \hbox{if } \ -0.125 \le \delta_i \le 0.08 \\
0.08,     & \hbox{if } \ \delta_i > 0.08 \\
\end{cases}
\end{equation}
Here, it is important to observe the upper limit of 8\% on $\theta_i$ {\it vs.} a lower limit of -12.5\%.

\medskip

\noindent
Step 3: Calculate 
\begin{equation}
\label{thetabar}
\bar\theta = \frac{1}{15} \sum_{i=1}^{15} \theta_i,
\end{equation}
the arithmetic mean of $\theta_1,\ldots,\theta_{15}$.  

\medskip

\noindent
Step 4: For the period March 15, 2007 -- March 15, 2008 the coupon rate, denoted by $R$, on the yield magnet note is determined as follows: 
\begin{equation}
\label{ymcouponrate}
R = \begin{cases}
0, & \hbox{if } \ \bar\theta < 0 \\
\min(\bar\theta,0.08) & \hbox{if } \ \bar\theta \ge 0 \\
\end{cases}
\end{equation}

\medskip

\noindent
Step 5: The coupon rate for March 15, 2008 -- March 15, 2011 will be no smaller than the rate for March 15, 2007 -- March 15, 2008.  Also, if any $\theta_i = 0.08$ on any D-date then that $\theta_i$ is fixed at $0.08$ for all later D-dates; this explains the phrase ``yield magnet.''
  
\bigskip

An advantage of yield magnet notes is that, by (\ref{ymcouponrate}), the coupon rate will be nonnegative for the period March 15, 2007 -- March 15, 2008 and will remain nonnegative for the full maturity period because of the ``yield magnet'' feature.   So, note-holders will lose no money because of fluctuations in stock prices.  

However, we also find a major disadvantage of the notes.  By (\ref{ymcouponrate}), we see that 
\begin{equation}
\label{ymcouponrateupperbound}
R \ \begin{cases}
= 0, & \hbox{if } \ \bar\theta < 0 \\
\le 0.08, & \hbox{if } \ \bar\theta \ge 0 
\end{cases}
\end{equation}
By applying to (\ref{ymcouponrateupperbound}) the Law of Total Expectation, we obtain an upper bound for the expected coupon rate:
\begin{equation}
\label{eymcouponrate}
E(R) \le 0.08 \cdot P(\bar\theta \ge 0).
\end{equation}

What we now deduce from this inequality would have been worrying to a note-holder in early 2007.  Stock prices had increased for years at that stage, so there was a low chance of further substantial increases in the 15 Titans' stock prices.  It was likely that most of the fifteen $\delta_i$ in (\ref{deltai}) were about to become substantially negative, the corresponding $\theta_i$ in (\ref{thetai}) then would equal -0.125, and so we would obtain $\bar\theta < 0$.  Hence, there was a low chance on March 15, 2007 that $\bar\theta$ would be positive over the next year.  If we suppose that $P(\bar\theta \ge 0) =  0.1$, or 10\%, then it follows from (\ref{eymcouponrate}) that 
$$
E(R) \le 0.08 \cdot 0.1 = 0.008
$$
Under the described scenario, the average note-holder would have obtained a coupon rate of {\it most} 0.08\%, or 8 basis points, over the period March 15, 2007 -- March 15, 2008.  Since higher rates were available in Federal Deposit Insurance Corporation (FDIC)-guaranteed bank accounts, such a coupon rate was insufficient to reward note-holders for the higher risk in a junior liability of an investment bank.  

We also find it disconcerting that the cutoff values in (\ref{thetai}) were chosen to be -12.5\% and 8\%.  According to that rule, the $\theta_i$ are skewed toward negative values, especially after early 2007; hence, there was a significantly high chance that $\bar\theta$ would be negative.  According to (\ref{ymcouponrate}), the coupon rate $R$ would be set to $0$, so a yield magnet note-holder effectively would be making an interest-free loan to the investment bank.  

By (\ref{thetai}) and (\ref{thetabar}), it follows that $-0.125 \le \bar\theta \le 0.08$; that is, the coupon rate will range between -12.5\% and 8\%.  If the stock prices of the 15 Titans each increased by 10\%, say, then note-holders would receive a return of 8\% and the bank would retain the remainder.  On the other hand, if the prices of the 15 Titans each declined by 10\% then holders will obtain a return of 0\%.  Although this strategy provides some protection for holders, the junior status of the notes places them in a grave position were the bank to default.  Hence, yield magnet note-holders were at a disadvantage relative to the bank regardless of whether stock prices advanced or declined.

\subsection*{Summary on yield magnet notes}
\label{yieldmagnetconclusions}

The owners of yield magnet notes will not know their coupon rates until the D-date, which is close to the payment date.  Retired investors, especially, many of whom are hard-pressed by currently low interest rates and heavily dependent on coupon payments to pay for living expenses, surely will have anxious moments as yield magnet payment dates draw closer.  Moreover, the complex nature of the coupon rate calculation indicates that a small investor likely will find it difficult even to ascertain personally that the reported calculation of the coupon rate has been done correctly.  

We are especially concerned by the upper limit of 8\% on the $\theta_i$ as opposed to a lower limit of -12.5\%.  This choice of cut-off limits is unsettling, and it is difficult to avoid the conclusion that those limits were selected to place note-holders at a disadvantage.  

Another issue is that the notes may complicate a buyer's income tax returns.  The prospectus states that ``comparable yield may be more than actual yield''on the notes, and further describes the notes as ``contingent payment debt instrument.''  These statements indicate that a potential buyer of the notes would be well-advised to consult an income-tax advisor before making an actual purchase.  

It is remarkable that the notes carried a fixed coupon rate of 5.5\% for the period March 15, 2006 - March 15, 2007.  Bearing in mind that a rate of 5.5\% was enormously higher than interest rates paid by ordinary FDIC-guaranteed banks accounts or money-market funds, we sense that the rate of 5.5\% simply was ``bait'' to entice clients who were desperate for higher yields.  

There is also the issue of whether there existed a {\it liquid} secondary market to service note-holders who needed to sell their notes before the redemption date.  We suspect that there was no such market, so that note-holders would be at the mercy of the banks, with resulting bids being significantly below the intrinsic value of the notes.  

Yield magnet notes were sold to investors as world stock markets neared record highs.  It would be interesting to carry out a post-mortem analysis to determine the resulting coupon rates since Fall 2008.  Bearing in mind the precipitous drop in the Titans Index from mid-2007 until early 2009, it is likely that yield magnet notes have yielded 0\% returns to note-holders for a lengthy period.  In that case, investors in yield magnet notes became unwitting sources of interest-free loans to investment banks.  

A detailed analysis of the probabilistic behavior of the coupon rate would be complicated.  Since the $\theta_i$ are correlated random variables then it would be difficult to estimate the probability that $\bar\theta > 0$, especially so when world equities had reached record territory and historical data were unlikely to be a useful guide to the future.  

\medskip

In conclusion, we find it difficult to understand why an investment adviser, in exercising {\it fiduciary care} of clients' funds, would have advised a client to purchase yield magnet notes during the period 2005--2007.

\section{Reverse exchangeable securities}
\label{reverseexchangeablesecurities}
\setcounter{equation}{0}

Reverse exchangeable securities are in one sense, the opposite of yield magnet notes.  Whereas yield magnet notes are based on a collection of stocks chosen from a global index, reverse exchangeable securities are based on a {\it single stock in a list chosen by the investment bank}.  This raises immediately a warning flag, for it is unlikely that such a list was chosen at random; indeed, a skeptical investor should assume immediately that the list was chosen to benefit the investment bank rather than the client.  

On that basis alone, a cautious investor should decline the reverse exchangeable securities.  Indeed, a skeptical investor surely should suspect that a financial company selling reverse exchangeable securities has chosen the single stock from a list in which it owns large holdings and is simply looking for ways to hedge its risk.  

Nevertheless, let us continue our analysis of these securities.  From a typical prospectus available at the SEC website, we find that reverse exchangeable securities are sold at a price of \$1,000; carry a maturity date of 1 year; and have a coupon rate of 12\%, paid in quarterly installments.  

The scheme for repaying the buyer at maturity is interesting.  If the stock has advanced then the buyer is paid \$1,000 in cash, so the return on investment is 12\%.  

If the stock has declined then, instead of repaying the initial purchase amount of \$1,000, the banker delivers to the buyer shares which were priced at \$1,000 at the start of the maturity period.  For example, suppose that the company's stock price had declined by 15\%, from \$1.00 to \$0.85; then, at the end of the maturity period, the banker delivers to the investor 1,000 shares, now worth \$850.  Therefore, the buyer's total return on the investment of \$1,000 is (12-15)\%, or -3\%, the difference between the coupon rate and the percentage decline in the stock price.  

Consequently, we obtain a general formula for $R$, the buyer's total return on capital.  Let $X$ denote the percentage change in the stock price; then 
\begin{equation}
\label{reverseexchangeablereturn}
R = 12 + \begin{cases}
0,   & \hbox{if } X > 0 \\
X, & \hbox{if } X \le 0
\end{cases}
\end{equation}

Applying the Law of Total Expectation to (\ref{reverseexchangeablereturn}), we obtain the following formula for the expected return on the reverse exchangeable securities: 
\begin{equation}
\label{reverseexchangeableexpect}
E(R) = 12 + E(X \, | \, X \le 0) \cdot P(X \le 0).
\end{equation}

Let us recall now that reverse exchangeable securities were sold after an extensive rise in worldwide equities prices.  Therefore, the average buyer of the securities should not expect to see continued increases in stock prices, and the probability is small, 5\% say, that a stock chosen at random from the banker's list will rise.  Then $P(X > 0) = 0.05$, so $P(X \le 0) = 0.95$.  If we also suppose that the stock prices in the banker's list behave similarly to the S\&P 500 index that declined by 42\% then we can suppose that $E(X | X \le 0) = -42$.  By (\ref{reverseexchangeableexpect}), we obtain 
$$
E(R) = 12 - (42 \cdot 0.95) = 12 - 39.9 = -27.9
$$
Here, the average owner of reverse exchangeable securities will experience a devastating loss of 27.9\% of their capital.

Suppose we were to adopt a more optimistic outlook, that there is a 10\% chance of the stock rising; so, $P(X > 0) = 0.1$ and $P(X \le 0) = 0.9$.  Assuming also that the average stock in the banker's list will fall only one-half as much as a broad index, i.e., $E(X | X \le 0) = -21$, then we obtain from (\ref{reverseexchangeableexpect}), 
$$
E(R) = 12 - (21 \cdot 0.9) = 12 - 18.9 = -6.9
$$
Again, the average owner of reverse exchangeable securities will experience substantial capital losses, {\it viz.}, 6.9\% of their capital.

Even under an extremely optimistic outlook, where we estimate that there is a 50\% chance of the stock rising, i.e., $P(X > 0) = 0.5$ and $P(X \le 0) = 0.5$, and where we also assume that the average stock in the banker's list will fall only one-quarter as much as a broad index, i.e., $E(X | X \le 0) = -10.5$, we obtain 
$$
E(R) = 12 - (10.5 \cdot 0.5) = 12 - 5.25 = 6.75,
$$
which is positive.  Under this rosy scenario, the average owner of reverse exchangeable securities will have received a positive return on capital; however, that return is not so large as to justify the enormous risks inherent in purchasing the securities.  

\subsection*{Summary on reverse exchangeable notes}
\label{reverseexchangeableconclusions}

Reverse exchangeable notes were sold to the public when stock markets neared record heights and there was a low chance that stocks would continue to rise significantly.  A skeptical investor should have assumed that $P(X \ge 0) \approx 0$.  Therefore, holders of these notes were doomed from the start to suffer substantial loss of capital.  

One the one hand, if the stock were to advance sharply then the bankers would pay the note-holder 12\% and retain any further profits.  On the other hand, if the stock were to decline then the notes allowed the bankers to put the depreciated stock to note-holders.  Such an arrangement was ideal for bankers who could unload stocks on unwitting note-holders and even to hedge call options on the stock.  

Any subsequent attempt by note-holders to sell the stock to limit their losses would simply depress the stock price further and enlarge their losses.  As large numbers of sell orders arrive at the stock exchanges, one can expect that the bankers would be tempted to sell the stock short (it would be nothing personal; it would simply be ``business'').  

That the coupon rate was set at 12\%, far higher than the existing near-5\% rate on 10-year U.S. Treasuries, should have been a major warning flag for note-buyers.  Such a wide difference between the two rates suggests that the notes were of junk-bond quality and therefore were unsuitable for unsophisticated investors.  In this regard, it would be interesting to determine the ratings assigned by ratings agencies to reverse exchangeable notes at the time of their creation.  The 12\% coupon rate also indicates that the quants who designed the notes were expecting stock indices to decline by far more than 12\%; as it turns out, they were correct in that expectation.

\medskip

To conclude, we find it difficult to understand why any investment adviser, in exercising {\it fiduciary care} of clients' funds, would have advised a client to purchase a reverse exchangeable note during the period 2006--2008.

\section{Principal-protected notes}
\label{principalprotectednotes}
\setcounter{equation}{0}

Another class of structured products that caused widespread losses were the ``guaranteed principal-protected'' notes.  Here, a skeptical investor should immediately ask two key questions:  ``Guaranteed by whom, or what?'' and ``Principal-protected by whom, and how?''  A paranoid\footnote{The sound advice of Grove \cite{grove} points out that ``only the paranoid survive.''} investor should detect immediately that the phrase ``guaranteed principal-protected'' contains redundancy in that ``guaranteed'' and ``principal-protected'' automatically imply each other; this observation alone should immediately heighten fears, leading a cautious investor to decline these notes.  

The typical ``fully principal-protected note with partial exposure to the S\&P 500 index'' is sold for a price of \$1,000, a maturity period of three years, and no coupon payments during that period.  Buyers may choose a ``participation rate,'' to be used to calculate the final return; in our analysis, we choose a participation rate of 80\%.  

As in (\ref{spreturn}) we calculate $I = (I_1-I_0)/I_0$, the percentage return of the index over the maturity period.  Then the return at maturity to the note-holder is 
\begin{equation}
\label{ratepppnote}
R = \begin{cases}
0, & \hbox{if } I \le 0 \\
0.8\, I, & \hbox{if } I > 0
\end{cases}
\end{equation}
The factor of 0.8 explains the ``participation rate of 80\%.''  

On applying the Law of Total Expectation to (\ref{ratepppnote}), we obtain 
$$
E(R) = 0.8 \, E(I \, | \, I > 0) \cdot P(I  > 0).
$$
Noting that these structured products were sold in large quantities in 2006-2008  after years of substantial increases in stock prices, a cautious investor should assume that the probability of further increases over a subsequent three-year period is small; that is, $P(I > 0)$ should be taken as small, at most 10\%, say.  Moreover, even if $I$ were to be positive over the maturity period, the investor should expect that $I$ will be small, so that $E(I \, | \, I>0)$ also should be taken to be small.  Even under a bullish outlook for the maturity period, with $E(I \, | \, I>0)$ assumed to be 25\%, we obtain from (\ref{ratepppnote}), 
$$
E(R) = 0.8 \cdot 0.1 \cdot 0.25 = 0.02
$$
Under this scenario, the average note-holder will receive a total return of 2\%.  Such an expected return clearly is insufficient compensation for the added risk of ``exposure'' to owning an unsecured security of a financial concern, especially so when FDIC-guaranteed certificates-of-deposit provided higher returns over the same period.  

We encourage the reader to calculate the expected return for (\ref{ratepppnote}) under alternative scenarios.  We have found that only highly optimistic, and hence naive, assumptions about the stock index lead generally to reasonable expected returns to note-holders.  

We cannot pass up an opportunity to mention the devilishly clever ``principal-protected absolute return barrier notes.''  An attentive reader is well-aware by now that the mere title of these products would cause us to decline them, but it will still be instructive to study them briefly.  We suppose that these notes are based on the S\&P 500 index, but readers may substitute another index of their choice.  As usual, we denote by $I_0$ and $I_1$ the closing levels of the S\&P 500 on the initial and final dates of the maturity period.  The prospectus stipulates an {\it upper barrier}, $U = 1.4 \, I_0$, and a {\it lower barrier}, $L = 0.6 \, I_0$, and then the note pays at maturity a return, 
$$
R = \begin{cases}
|I|, & \hbox{if the index remains between } U \hbox{ and } L \hbox{ \it for the entire maturity period} \\
0, & \hbox{otherwise}
\end{cases}
$$
Note-holders obviously must be hoping that the index remains between $U$ and $L$ for the entire maturity period.  But given that the S\&P 500 index can be notoriously volatile over short time periods, such hopes are likely to be dashed.  

By the Law of Total Expectation, 
\begin{align*}
E(R) = \ & E\big(|I| \, \big| \, \hbox{The index remains between } U \hbox{ and } L \hbox{ for the entire maturity period}\big) \\
& \ \cdot P\big(\hbox{The index remains between } U \hbox{ and } L \hbox{ for the entire maturity period}\big).
\end{align*}
We can estimate both terms in this formula using historical data on the index, however such data in 2006-2007 would have predicted the index inaccurately over the ensuing three years.  Assuming that the underlying issuer remained solvent, we suspect that absolute-value barrier return note-holders received 0\% returns on average.  In any case, our brief analysis would have led us to decline these notes in 2006-2007.  

Deng, et al. \cite{deng} appear to have found other flaws in principal protected absolute return barrier notes; however, they used mathematical techniques which are more advanced than we are willing to apply in the study of any financial product.  We recall here a remark of Graham \cite{graham}, p. 321 who commented that, in the valuation of common stocks, whenever ``calculus is brought in, or higher algebra, you could take it as a warning signal that the operator was trying to substitute theory for experience, and usually also to give to speculation the deceptive guise of investment.''  
Going beyond Graham's sage advice, {\it we urge unsophisticated and sophisticated investors to view askance any financial product whose valuation requires the use of calculus or higher algebra}.

\subsection*{Summary on principal-protected notes}
\label{principalprotectednotesconclusions}

On reading various prospectuses for ``guaranteed principal-protected'' notes available on the  SEC website, we can dissect the inner workings of those products.  The analysis that we have done in the previous sections of this paper can be repeated for these notes, and the conclusions are the same: Note-holders in 2006-2007 were doomed to suffer massive losses.  Many of these notes were junior liabilities of the issuing banks, so when the financial crisis in the U.S. worsened in late 2008 and some banks became insolvent, the true nature of the ``guaranteed principal-protected'' rubric became clear, viz., they were neither guaranteed nor principal-protected (Morgenson \cite{morgenson}).

There were also ``guaranteed principal-protected notes'' which offered buyers ``partial exposure''\footnote{In the old days, people who exposed themselves to the elements were viewed either as lacking good manners, behaving criminally, or mentally insane.  By contrast, investors today do not hesitate to expose themselves to the elements of high-risk ventures for which they have no understanding.} to various indices based on equities, commodities, interest rates, or currency exchange rates.\footnote{Warning: Speculation in interest rates or currencies are, for small investors, simply sheer madness!}  Our method of analysis will show again that those who bought these products in 2006-2008 were doomed to suffer huge losses.  

In conclusion, we find it difficult to understand why an investment adviser, in exercising {\it fiduciary care} of clients' funds, would have advised a client to purchase principal-protected notes during the period 2006--2008.

\section{Other structured products}
\label{otherproducts}
\setcounter{equation}{0}

A panoply of structured financial products were invented by the quants and sold to the public by investment banks during the past decade.  The list of these products is so lengthy as to be bewildering, but they all seem to have several features in common with the four types which we have analyzed here:

\smallskip
\noindent
1.  They all have ``interesting'' names, many with stylish acronyms.  

Examples are ``stock participation accreting redemption quarterly-pay securities'' (SPARQS), ``dual directional contingent buffered equity notes,'' ``buffered underlying securities'' (BUyS), and ``knock-out reverse convertible notes.''\footnote{Even unsophisticated investors should instinctively know better than to buy a financial product whose name contains the phrase ``knock-out.''  Many of them, unfortunately, did not know better and indeed were KO'd ``in the first round'' \cite{pageperry}.}  And there are the ``protected notes'' (which came in a staggering list of variations including: basket cliquet, Everest, rainbow, Atlas, Himalaya, pulsar, Antiplano, Neptune, and many more), digital accrual notes, barrier accrual floating notes, ratchet notes, currency swaps, and other exotica (Bateson \cite{bateson}).  

\smallskip
\noindent
2.  They are all based on complicated mathematical formulas.  

These formulas are so complicated that not even the quants appear to understand the full implications of the Frankensteins that they invented.  Then the stage for disaster is set when the underlying assumptions are violated by real-world events which were not envisaged by the quants.  

\smallskip
\noindent
3.  They collectively have caused unsophisticated investors billions of dollars in losses, generating thousands of lawsuits.  

Even some formerly-sophisticated, prominent market participants have lost large amounts on structured products.  An example is J.P. Morgan Chase which admitted multibillion-dollar losses recently from trading financial derivatives (Silver-Greenberg \cite{silvergreenberg}).  It is ironic that the largest innovator and issuer of derivatives itself has lost vast amounts on derivatives.  

\smallskip
\noindent
4.  They encourage investors to hope for outcomes which are too good to be true and with no thought to the consequences of adverse events.  

Examples of these are ``return-enhanced'' optimization securities, which pay investors a multiple of an index's gain but provide no protection against a decline.  

\medskip

Without stringent regulatory oversight that includes probabilistic analyses, the quants will continue to invent new products, for there is great pressure from the bankers to do so.  However, we expect that the new products will rarely benefit clients, and investors would do well to note the superbly incisive comment of Morgenson \cite{morgenson}: ``Add these securities to the growing pile of Wall Street inventions that benefit $\ldots$ wait for it, wait for it $\ldots$ Wall Street.''

\section{Conclusions, and implications for U.S. Treasuries}
\label{conclusions}
\setcounter{equation}{0}

In analyzing the four structured products, we did not study the implications of income taxes or sales fees arising from the purchase of those products.  Had we done so, the conclusions would be even worse for unsophisticated investors.  Few such investors are knowledgeable of the income-tax treatment of capital gains or losses on derivative securities, so they will need professional tax advice, which is likely to be expensive.  

As for fees, it was noted in \cite{barnum} that some banks were charging ``fees on some structured notes that equal or exceed the securities' highest possible yield'', a practice which brought to mind P. T. Barnum's ``sucker born every minute'' dictum.  

There is, ultimately, a dark side to the invention of derivative securities and, more broadly, to financial innovation.  Henderson and Pearson \cite{hendersonpearson} remarked that financial institutions can ``mislead investors to value the new instruments more highly than they would if they understood financial markets and correctly evaluated information about probabilities of future events.''  This comment is amplified by the probabilistic calculations that we have now done.  

This dark side manifests itself even in the naming of structured products.  Given that return optimization securities were based on Markowitz's solution to the ``risk--return optimization'' problem, it would have been natural for those products to be named ``risk--return optimization securities.''  However, sales departments are likely to argue that the word ``risk'' would scare customers; and legal departments would argue that the word ``risk'' would increase their legal liability when the products crashed.  So, the name was reduced to ``return optimization securities.''  And down went the clients with their {\it negative}-return optimization securities.  

Steinsaltz \cite{steinsaltz} noted that civil engineers' code of ethics enjoins them from ``finding bridges prone to collapse so they can cash in buying~insurance on them'', ``boasting of how their elite training and inherent brilliance enabled them to hoodwink the inspectors into approving shoddy plans'', or not ``examining painstakingly what will happen in case of an accident, a fracture or conflagration''.  By contrast, quants collectively appear to have no such qualms when designing structured products.  Therefore, the name ``financial {\it engineer}'' is a sham.  

In surveying the grief suffered by small investors worldwide who lost their investments in structured products we urge the quants, {\it especially those in academia}, to reflect on whether they bear any ethical responsibility for these events.  

Nevertheless, the responsibility for losses due to structured products must be placed ultimately on the shoulders of buyers; here, it is easy enough to mention the old maxim, ``{\it caveat emptor}.''  Moreover, we have used the term ``investor'' loosely in this paper, but the buyers of structured products actually were speculators, and they apparently were unaware of the distinction between the terms ``investor'' and ``speculator'' (Graham \cite{graham}, p. 18).  As a practical matter, we urge them to read Graham and Dodd \cite{grahamdodd}, p. 66 to relearn the importance of caution in purchasing fixed-income securities:  

\begin{quote}
\begin{small}
``{\bf Major Emphasis on Avoidance of Loss.}{$\boldsymbol{-}$}Our primary conception of the bond as a commitment with limited return leads us to another important viewpoint toward bond investment. Since the chief emphasis must be placed on avoidance of loss, bond selection is primarily a negative art. It is a process of exclusion and rejection, rather than of search and acceptance. In this respect the contrast with common-stock selection is fundamental in character. The prospective buyer of a given common stock is influenced more or less equally by the desire to avoid loss and the desire to make a profit. The penalty for mistakenly rejecting the issue may conceivably be as great as that for mistakenly accepting it. But an investor may reject any number of good bonds with virtually no penalty at all, provided he does not eventually accept an unsound issue. Hence, broadly speaking, there is no such thing as being unduly captious or exacting in the purchase of fixed-value investments. The observation that Walter Bagehot addressed to commercial bankers is equally applicable to the selection of investment bonds. `If there is a difficulty or a doubt the security should be declined.' ''(\cite{bagehot}, p. 245).
\end{small}
\end{quote}

Although much of the blame for the debacle of structured products lies with the buyers, bankers, and quants, there is a fourth party which deserves even more blame: the U.S. Federal Reserve.  For the past decade, the Fed has kept interest rates so low that it has placed fixed-income investors on the horns of a dilemma.  According to Zweig \cite{zweig}, a prominent value investor commented that, by holding interest rates at zero, the Fed is ``giving bad advice'' to investors to swap their safe, 0\%-interest bank accounts for unsafe securities having higher risk than perceived and ``basically tricking the population into going long on just about every kind of security except cash, at the price of almost certainly not getting an adequate return for the risks they are running.''  

Moreover, the low interest rates have had two particularly pernicious consequences.  First, the low rates have enabled financial firms to indulge in unprecedented speculative activities, some of which have had negative outcomes that have prolonged a general lack of confidence in the financial markets, extended the on-going financial crisis, and weakened the economy.  

Second, the continuing lack of confidence and weakened economy have heightened fears among investors, causing many of them to purchase U.S. Treasury securities at negative real-interest rates.  This has helped to increase the prices of U.S. Treasuries to record, bubble-like levels, and we truly fear for the day when this trend reverses.  

We conjecture that the current financial crisis will last until at least 2018.

\section{Postscript}
\label{postscript}

On March 14, 2012 the law firm of Dimond, Kaplan \& Rothstein \href{http://www.dkrpa.com/blog/2012/03/lehman-brothers-investors-still-have-time-to-pursue-finra-claims.shtml}{\textcolor{blue}{issued a report}} on a formerly-sophisticated investment bank that emerged recently from bankruptcy reorganization.  The report stated that investors in the bank's principal protected notes, partial protection notes, step-up callable notes, return optimization securities, and absolute return barrier notes likely will recover through the bankruptcy reorganization only about 20\% of their investment losses.  ``For many investors, the only way to recover the remaining 80\% of their investments losses is through a FINRA arbitration claim'', the report continued.  Such investors were urged to file claims in a timely manner.

\addcontentsline{toc}{section}{References} 

\bibliographystyle{ims}

\begin{thebibliography}{00}

\parskip=3.9pt

\bibitem{bagehot}
W. Bagehot. {\sl Lombard Street}.  C. Scribner's Sons, New York, 1892.

\bibitem{bateson}
R. D. Bateson. {\sl Financial Derivative Investments: An Introduction to Structured Products}.  Imperial College Press, London, 2011.

\bibitem{barnum}
Bloomberg News. \href{http://www.investmentnews.com/article/20100624/FREE/100629938}{\textcolor{blue}{Paging P. T. Barnum: Fees on some broker-sold notes exceed}} \href{http://www.investmentnews.com/article/20100624/FREE/100629938#}{\textcolor{blue}{possible return}}.  June 24, 2010.

\bibitem{deng}
G. Deng, I Guedj, J. Mallett, and C. McCann.  The anatomy of principal protected absolute return barrier notes.  {\it The Journal of Derivatives}, {\bf 19} (2011), 61--70.

\bibitem{grahamdodd}
B. Graham and D. L. Dodd.  {\sl Security Analysis}. McGraw-Hill, New York, 1934. 

\bibitem{graham}
B. Graham.  {\sl The Intelligent Investor: A Book of Practical Counsel}.  4th revised edition,   
Harper, 1973.

\bibitem{grove}
A. S. Grove.  {\sl Only the Paranoid Survive}.  Doubleday, New York, 1999.

\bibitem{hendersonpearson}
B. J. Henderson and N. D. Pearson.  The dark side of financial innovation.  {\it Journal of Financial Economics}, {\bf 100} (2011), 227--247.

\bibitem{markowitz}
H. Markowitz.  The optimization of a quadratic function subject to linear constraints.  {\it Naval Research Logistics Quarterly}, {\bf 3} (1956) 111--133.

\bibitem{morgenson}
G. Morgenson.  \href{http://www.nytimes.com/2010/05/23/business/23gret.html}{\textcolor{blue}{Fair game: `100\% protected' isn't as safe as it sounds}}.  {\it The New York Times}, March 22, 2010, p. BU1.

\bibitem{pageperry}
Page Perry, LLC.  \href{http://www.investmentfraudlawyerblog.com/2010/05/jp_morgan_reverse_convertibles.html}{\textcolor{blue}{JP Morgan reverse convertibles ``knock out investors in the first}} \href{http://www.investmentfraudlawyerblog.com/2010/05/jp_morgan_reverse_convertibles.html}{\textcolor{blue}{round}''}.  May 18, 2010.

\bibitem{papini}
J. Papini.  \href{http://online.wsj.com/article/SB10001424052748703735004574576260110956526.html}{\textcolor{blue}{Investor wins Lehman note arbitration}}.  {\it The Wall Street Journal}, December 5--6, 2009, p. B.3.

\bibitem{richardsbgsu}
D. St. P. Richards.  \href{http://www.stat.psu.edu/~richards/talks/}{\textcolor{blue}{2006 Bowling Green Lectures}}.

\bibitem{richardshundal} 
D. St. P. Richards and H. Hundal.  \href{http://www.stat.psu.edu/~richards/papers/cpdo.pdf}{\textcolor{blue}{Constant proportion debt obligations, Zeno's}} \href{http://www.stat.psu.edu/~richards/papers/cpdo.pdf}{\textcolor{blue}{paradox, and the spectacular financial crisis of 2008}}.  In: {\sl Proceedings of the Eubank Conferences on Real World Markets}, Rice University, March 23--24, 2009 (forthcoming).

\bibitem{ross}
S. Ross.  {\sl A First Course in Probability}, 8th edition.  Pearson, Upper Saddle River, NJ, 2010.
  
\bibitem{silvergreenberg}
J. Silver-Greenberg.  \href{http://dealbook.nytimes.com/2012/06/12/jpmorgan-chief-says-huge-trading-loss-was-isolated-event/}{\textcolor{blue}{JPMorgan chief says huge trading loss was ``isolated event''}}.  New York Times, June 12, 2012.  

\bibitem{steinsaltz}
D. Steinsaltz.  \href{http://www.ams.org/notices/201105/rtx110500699p.pdf}{\textcolor{blue}{The value of nothing: A review of {\sl The Quants}}} [book review, Crown Business, New York, 2010].  {\it Notices of the American Mathematical Society} {\bf 58} (2011), 699--704.

\bibitem{zweig}
J. Zweig.  \href{http://online.wsj.com/article/SB10001424052748704167704575258442772338282.html}{\textcolor{blue}{Legendary investor is more worried than ever}}.  {\it The Wall Street Journal}, May 22, 2010.

\end{thebibliography}

\end{document}